%% file: main.tex
\documentclass[letterpaper]{article}
\usepackage{aaai25} 
\usepackage{times} 
\usepackage{helvet} 
\usepackage{courier} 
\usepackage[hyphens]{url} 
\usepackage{graphicx} 
\usepackage{graphicx} 
\usepackage{natbib} 
\usepackage{caption} 
\title{The Case for ``Thick Evaluations" of Cultural Representation  in  AI}

\author {
    Rida Qadri,
    Mark Díaz,
    Ding Wang,
    Michael Madaio
}
\affiliations {
Google Research \\
    ridaqadri@google.com, markdiaz@google.com, drdw@google.com, madaiom@google.com
}

\begin{document}

\maketitle
\begin{abstract}
Generative AI model outputs have been increasingly evaluated for their (in)ability to represent non-Western cultures.  We argue that these evaluations often operate through reductive ideals of representation, abstracted from how people define their own representation and neglecting the inherently interpretive and contextual  nature of cultural representation. In contrast to these ‘thin’ evaluations, we introduce the idea of \textit{`thick evaluations'}: a more granular, situated, and discursive measurement framework for evaluating representations of social worlds in AI outputs, steeped in communities' own understandings of representation. We develop this evaluation framework through workshops in South Asia, by studying the `thick' ways in which people interpret and assign meaning to AI-generated images of their own cultures. We introduce practices for thicker evaluations of representation that expand the understanding of representation underpinning AI evaluations and by co-constructing metrics with communities, bringing measurement in line with the  experiences of communities on the ground.\looseness=-1
\end{abstract}

\section{Introduction} \label{introduction}
\input{Sections/01_Introduction}

\section{Related Work} \label{related_work}
\input{Sections/02_Related_Work}

\section{Methods} \label{methods}

\input{Sections/03_Methods}
\section{Findings}
\input{Sections/04_FINDINGS_new}

\section{Discussion} \label{discussion}
\input{Sections/05_Discussion}

\section{Limitations and Future Work}
\input{Sections/06_Limitations}
\section{Conclusion}
\input{Sections/07_Conclusion}


\bibliography{acmart.bib}

\input{Sections/08_Appendix}

\end{document}

%% file: Sections/01_Introduction.tex
Generative AI (genAI) models, despite their popularity, have been shown to fail at inclusively representing different cultures in generated outputs---including images \cite{qadri2023,Mim2024,mack2024they} and text \cite{wang-etal-2024-countries,pawar2024surveyculturalawarenesslanguage,singh2024globalmmluunderstandingaddressing,myung2025blendbenchmarkllmseveryday, khanuja2021murilmultilingualrepresentationsindian}---much like failures of representation in other AI systems \cite[e.g.,][]{katzman2023taxonomizing,chien2024beyond,wang2024measuring, kay2015unequal, shelby2023sociotechnical,crawford2017trouble}.\looseness=-1

To address these gaps, prior work has sought to evaluate the cultural representations in AI-generated output, but with few exceptions \cite[e.g.,][]{ghosh2024caste, qadri2023}, mostly through quantified, metricized approaches to representation such as statistical similarities and benchmark-style scoring \cite{zhang2024partiality, li2024culture}. The use of these methods presumes that representation is an objective construct with a quantifiable, definitive ground truth that outputs can be compared against \cite[e.g.,][]{kannen2024beyond, zhang2024partiality} \cite[for a critique of ground truth, see][]{muller2021designing}. Given limitations of computational methods, representation is reduced to evaluations of basic recognition or factual generation of artifacts. When human feedback on representation is sought, it is often solicited through narrow, constrained, quantitative scales from crowdworkers who may not have the lived experiences to evaluate nuances of content's varied meanings to different social groups \cite{diaz2022crowdworksheets}.

However, this approach to measuring representation is in contravention to decades of scholarship in the social sciences that emphasizes the subjective nature of representation, where judgments about representation in visual media are constructed in conversation with the viewer's lived experiences  and the broader context within which  an image is viewed and published.  \cite{hall1989cultural,hall1997representation}.  In addition, the categories many AI researchers use for evaluations are often abstracted from the experiences of communities, constructed by AI researchers without engaging with communities to understand what axes of representation might be salient in their contexts. For instance, while many representational evaluations focus on skin tone, skin tone might not be a salient category of differentiation for social groups in many cultures. 
Thus, while  existing approaches to evaluating representation in AI/ML might be useful for evaluating the accuracy of visual depictions of the physical world, they may not allow us to engage with communities' diverse desires for  representations of their \textit{social worlds}.\looseness=-1 

  As generative image models increasingly are used to represent social worlds, what new evaluation approaches are necessary to meaningfully account for the diverse ways that people interpret and evaluate cultural representation in AI? 
\looseness=-1 

 We argue that effectively evaluating cultural representation in AI images requires \textbf{`thick evaluations'}---a more granular, situated, discursively constructed approach to measurement, steeped in communities' own understanding of appropriate cultural representation. 
 Our analysis draws on the `thick vs thin' dichotomy introduced by philosopher Gilbert \citet{rylethinking}, later taken on by Clifford \citet{geertz2008thick} and others \cite{ortner1995resistance,love,jackson2013thin, riles2000network,nelson2023facct,nelson2023oxford} to characterize
 descriptions of social worlds. Ryle's classic example of a wink illustrates this distinction: a `thin description' merely describes the physical act of closing one eye by focusing on observable details, while a `thick description' unpacks the social meaning of the act and its significance as a signal to interlocutors.   
Seen through this framework, emerging methods for evaluating cultural representation in AI images involve `thin evaluations'; i.e., suited to evaluating the observable aspects of the physical world contained in AI-generated images, but not necessarily the social signals embedded in those physical depictions.\looseness=-1  

To develop a `thick evaluation' approach for cultural representation in AI images, we turned to the communities represented in images we seek to evaluate, recognizing their expertise and stakes in visual representations of their cultures. Given prior work on AI's failures of representation for South Asian cultures \cite[e.g.,][]{qadri2023,ghosh2024generative}, we conducted workshops with 37 participants in 3 South Asian countries to study the culturally-situated ways people assign meaning to, interpret, and evaluate the representation of their cultures in AI-generated images. We find that 1) people evaluate representation of social worlds not just through a singular category such as accuracy, but through multi-dimensional, fine-grained axes; 2) people's goals for evaluating cultural representation are situated in their social context,  negotiated through dialogue with others and  in response to broader societal discourse about their cultures; and 3) people use situated social knowledge of and experiences with social worlds to evaluate varying social meanings of images.\looseness=-1 

These findings demonstrate that thin evaluations  alone cannot measure thick constructs like cultural representation, and they raise critical questions about the adequacy of existing evaluation paradigms for measuring cultural representation in AI images. As new methods and standards are being created for AI evaluations, this is a crucial moment to call for a thicker AI evaluation practice that  reflexively interrogates  epistemological underpinnings of AI evaluation practices, fundamentally rethinks whose expertise is included in the evaluation of  AI systems, and opens up space for more interpretive qualitative evaluation methods. We provide pathways for such thickness by showing how AI researchers can  interrogate the construct of cultural representation and the ways that that unobservable construct is operationalized in evaluation methods \cite[cf.][]{jacobs2021measurement}. We provide empirical data on varied categories of representation people might evaluate images for, which shows the need for co-constructing evaluation methods with members of communities whose culture is being represented, to bring measurement in line with their experiences of AI images.
We thus help AI practice move towards an ecosystem of both `thick' and `thin' evaluations, encouraging congruence between the construct being evaluated and the evaluation methods used. As AI technology is entering into the space of cultural production, our work suggests the need to develop  new, thicker forms of evaluations of cultural representation, to better reflect how people consume and interpret images of their cultures.\looseness=-1

%% file: Sections/02_Related_Work.tex
\subsection{Evaluating Cultural Representation in AI} 
\label{positivism}

In order to understand the nature of AI models' failures of representation \cite[e.g.,][]{qadri2023,mack2024they}, researchers have developed methods for evaluating representational failures (e.g., stereotyping) of AI systems \cite[e.g.,][]{katzman2023taxonomizing,chien2024beyond,wang2024measuring, kay2015unequal, shelby2023sociotechnical,crawford2017trouble,harvey2024gaps}. Primarily, these evaluations have focused on  language models \cite[e.g.,][]{blodgett2021stereotyping,hosseini2023empirical,abbasi2019fairness,katzman2023taxonomizing,chien2024beyond,wang2024measuring, wolfe2024representation}, but they have also begun to include image models. For instance, emerging approaches for evaluating representation in images include developing quantified benchmark scores for goodness of representation (e.g., diversity \cite{kannen2024beyond}), via methods like creating statistical similarity between generated images and reference images \cite{zhang2024partiality,ghosh2024caste}, calculating correlations in keywords \cite{li2024culture}, measuring model frequency of generating stereotypical and offensive images of nationality groups \cite{jha2024visage}, and calculating differences in frequency between the most and least common identities referenced in a set of model outputs \cite{lahoti2023improving}. Human feedback has been sought via, for instance, using anonymized crowdworkers to score cultural biases in images \cite{basu2023inspecting, ghosh2024don}. For representation in images, these measures often focus on visual demographic diversity characteristics---for instance,
\citet{Cho22} use automated skin tone and gender presentation classifiers to evaluate diversity of generated images. \looseness=-1

However, this line of work follows a technosolutionist, positivist \cite[][]{crabtree2024h} conception of representation; i.e., treating representation as something objective, stable across time and contexts, and quantifiable. For instance, using skin tone to measure diversity presumes skin tone is a useful proxy for race, and race is a universal category of differentiation, which may not be appropriate in all global cultures being represented.   In contrast, researchers at the intersection of HCI and responsible AI have conducted qualitative evaluations of representation in genAI image models, identifying failures of representation through more participatory and community-centered methods, with evaluators who are from the same culture as the images they are evaluating  \cite{qadri2023,  Mim2024,ghosh2024don}. However, while these efforts have demonstrated the value of qualitative approaches for evaluating representation in generative image models, they do not examine what is meant by representation as a construct, nor do they critically interrogate what the participants' process of interpretive meaning-making suggests for methods for evaluating cultural representation in AI more broadly—both of which we focus on in this paper.\looseness=-1

Recent research has shed new light on the complex process of measuring unobservable social constructs such as fairness, as well as the potential harms of reductive measurement approaches. For instance, \citet{selbst2019fairness} critique the definitions of fairness that lead to misleading abstractions in its evaluation. Recognizing the contested nature of the construct of fairness, \citet{Smith-design-fairness} argue for  bringing in stakeholders' expertise for more meaningful and contextually grounded evaluations. Scholars like \citet{jacobs2021measurement} advocate for utilizing approaches to \textit{measurement modeling} from the social sciences to bridge the gap between unobservable constructs and their operationalization in evaluations of fairness in algorithmic systems (and for evaluation of generative AI \cite[e.g.,][]{wallach2025position,weidinger2025toward})---a framework we draw on in this paper through our discussion of the construct of cultural representation and how it is operationalized in evaluations.

\citet{jacobs2021measurement} argue that measurement modeling begins with a clear and robust articulation of the construct being evaluated---in their paper, fairness, and in our paper, cultural representation. However, in prior work evaluating cultural representation in text-to-image models \cite[e.g.,][]{zhang2024partiality,kannen2024beyond,jha2024visage,lahoti2023improving,basu2023inspecting}, the construct of representation is often not interrogated or only implicitly defined. Although often left implicit, prior work operationalizes representation in various ways, as lack of bias \cite{wan2024survey}, as geo-cultural similarity \cite{basu2023inspecting}, and as imbalance \cite{su2009handling}. More broadly than for cultural representation in text-to-image models, \citet{chasalow2021representativeness} argue that researchers often fail to make explicit how they conceptualize `representative' or `representation.' Such various ways of operationalizing the same construct may lead to evaluations that use the same term, but optimize for disparate goals (e.g., cultural awareness, cultural diversity, avoiding ``cultural bias'' \cite{li2024culture}, etc), making it difficult to compare the effectiveness of different approaches.\looseness=-1
 
\subsection{Situated Conceptualizations of Representation} 

Decades of scholarship from the humanities have argued that representation in media, particularly visual media, cannot be defined through positivist approaches \cite[e.g.,][]{hall1989cultural,hall1997representation,barthes1999rhetoric}. They argue that representation is not a static goal that can be objectively and quantitatively evaluated and achieved, but is instead an ongoing interpretive act whose meaning is contested and negotiated. In media studies, visual media, such as paintings, photography, or digital art, do not represent an objective reality but instead communicate multiple shades of meaning about the world  \cite{hall1997representation}. That is, images have both denotative meanings (i.e., the literal subject of the image) as well as connotative meanings (i.e., the emotional subtext or symbolic meaning) the image was designed to convey---or the meaning it evokes in the viewers, which may be different than the intended meaning of the image \cite{hall1997representation, barthes1999rhetoric}. 

As a result, for media studies scholars, representation in visual media is a process of meaning-making, not a one-to-one depiction of the world as it is---and thus representation is a site for struggle \textit{over} meaning \cite{hall1989cultural,hall1997representation,desai2000imaging}. In this way, even a photograph of an event does not objectively convey a singular true meaning \cite{hall1997representation}, but instead conveys a ``positional truth'' \cite{abu1991before} (or multiple such truths, given the interpretive nature of images), shaped by the creators' and viewers' cultural and historical context. Thus, visual understanding itself relies on a slippery relationship between words, images, and the concepts signified by those images \cite{barthes1999rhetoric}.

As media studies scholars remind us, representation in public culture is a ``zone of contestation'' \cite{desai2000imaging}, characterized by contests over which peoples and perspectives are made visible, how they are portrayed (and by whom), contests which are shaped by who has the power to control narratives of representation. Thus, in this paper, we explore how people evaluate representation in AI-generated images and what this representation means to them in their culture.   

\subsection{Thick vs Thin Modes of Understanding the Social}  
The `thick/thin' dichotomy, originating in Gilbert Ryle's philosophical work (\citeyear{rylethinking}), has become a cornerstone of anthropological and social scientific inquiry.  He argued that understanding human behavior requires moving beyond mere observation of physical details to interpretation, recognizing the ``many-layered sandwich'' of meaning embedded within even seemingly simple actions \cite{rylethinking}.  This concept was further elaborated upon by Clifford \citet{geertz2008thick}, who emphasized the importance of thick description in ethnographic research. Geertz argued that researchers must not only describe actions but also interpret them within their social or cultural context. A wink, for instance, could signify a conspiratorial gesture, a flirt, a parody, or a mere rehearsal, each carrying distinct meanings depending on the social context. Sherry Ortner, in a chapter on ``thick resistance,'' highlighted the limitations of thin approaches in capturing the complexities of social life, which require the \textit{``richness, texture, and detail,''}  of thickness  \cite{ortner1995resistance,ortner1997thick}. She argued that thin descriptions often overlook the \textit{``internal politics of dominated groups, thin on the cultural richness of those groups, thin on the subjectivity---the intentions, desires, fears, projects---of the actors engaged in these dramas''} \cite{ortner1995resistance}. Ortner advocated for thick descriptions that delve into the \textit{``production, circulation, and consumption''} of cultural practices, recognizing their dynamic interplay with social and political realities. More recently, Alondra Nelson has called for ``thick alignment'' when aligning AI systems with human values \cite{nelson2023facct,nelson2023oxford}. While the value of thickness in capturing the nuances of social phenomena has been widely acknowledged, scholars have also recognized the importance of thinness in discerning patterns and broader social structures. Annelise Riles \citet{riles2000network}, in her work on multi-sited ethnography, explored the challenges of achieving thick description when studying global phenomena dispersed across diverse cultures, calling attention to \textit{``the limits of thickness as a disciplinary trope''} and to pay more attention to what the use of ``thin composition'' could yield. Anthropologist John Jackson similarly emphasized the value of thin description in providing a \textit{``baseline empiricism''} and a starting point for social investigation \cite{jackson2013thin}. Thus, while thick and thin both have their place in social analysis, each one can only get you so far. 
In this paper, we draw on the thick/thin dichotomy to provide a framework for navigating the complexities of social inquiry when considering multiple approaches to evaluating cultural representation in AI images.\looseness=-1

%% file: Sections/03_Methods.tex
In this research, we treat the evaluation process as an object of study itself through community-centered, qualitative methods. These methodological choices respond to increasing calls for broader participation in AI development and design \cite{delgado2023participatory,birhane2022power}.
We engaged with participants from three South Asian countries (Sri Lanka, Pakistan, and India) through a series of individual online evaluations and collective in-person workshops. These methods were chosen to allow us to move beyond binary evaluation metrics like thumbs up/down \cite[cf.][]{collins2024beyond}; we instead sought to capture in-depth qualitative reflections and discussions that participants have as they interpret and evaluate AI-generated images. Given the discursive and collective nature of representation in media, we chose group workshops instead of individual interviews. To preserve participants' perspectives, we included an individual evaluation task prior to the workshops. Additionally, to ensure the annotation process reflected local cultural norms—including platform choice, framing of questions, and participant composition—we collaborated with local research partners in each region.\looseness=-1   

\subsection{Participants and Sites}

We chose our research sites for two main reasons. First, we aim to contribute to a growing body of scholarship that seeks to expand the evaluation of AI beyond Western-centric perspectives \cite{kak2020global, mohamed2020decolonial, sambasivan2021re, qadri2023,ghosh2024generative}. Second, we intend to move beyond the common ``USA vs. India'' dichotomy often found in comparative analyses including the Global South \cite[e.g.,][]{raman2008portrayals, khairullah2009cross}. To this end, we focused on three South Asian countries: Sri Lanka, Pakistan, and India. They were selected due to the shared cultural histories, while also recognizing the vast diversity within South Asia.

We recruited participants through collaborations with local research partners deeply embedded in their communities. Our partners in Sri Lanka and Pakistan were university researchers with extensive research experience with marginalized local communities, while our partner in India was a research manager experienced in conducting workshops and focus groups. Recruitment began with emails to targeted lists, including local institutions, previous research participants that worked with those partners, and the partners' professional networks. Interested individuals completed a screening form detailing demographics and background.

Rather than attempting to exhaustively represent any single culture, our aim was to capture nuanced diversity and similarities about how people from different cultures interpret and assign meaning to their cultural representation. Thus our selection criteria took a purposive sampling approach \cite{blandford2016qualitative}, aiming to identify a diverse set of individuals with various intersecting identities relevant to the local context. We were not prescriptive regarding the exact numbers and form of diversity in our sample, but instead we relied on our partners to guide us towards axes of marginalization and social differentiation that might influence socio-cultural experiences in the country. These led to a sample that was highly contextualized in its diversity but not necessarily exhaustive of all forms of diversities. Minority and diversity here are interpreted in the context of the respective country. For instance, Hinduism is not a minority religion in India, but it is in Sri Lanka and Pakistan, while being Muslim is not a minority in Pakistan, but it is in Sri Lanka. 15 participants per country were selected. Ultimately, 11 participants each from Sri Lanka and Pakistan and 15 from India (spread across two workshops) took part. Participants' backgrounds are detailed in the Appendix.\looseness=-1

\subsection{Study Activities}
Our study comprised three parts: a pre-workshop \textbf{survey} to elicit prompts for culturally relevant images, a pre-workshop \textbf{evaluation task} to evaluate those images, and a \textbf{workshop} to discuss cultural representation in AI images. 

\textbf{Pre-workshop surveys.}
We asked each participant to create prompts for AI image generation, drawing on their unique cultural experiences by mixing  salient identity markers (e.g., ``Punjabi woman'') with a regional landmark or cultural event, resulting in prompts such as ``a group of Punjabi women in front of Wazir Khan Mosque in Lahore.'' This fostered diverse and culturally-relevant prompts and enabled participants to be experts in evaluating the resulting images.\looseness=-1 

\textbf{Individual evaluation task.}
We wanted to give each participant space to individually evaluate the images about their cultures generated by their prompts, before joining the workshop. Thus, after we gathered the prompts from the participants, one of the authors generated images based on each participant's prompts and compiled a personal collection of prompt and images for each person. Participants were invited to reflect on each image and leave commentary on whether the image was a good representation of their cultural experience. We purposefully did not define ``good'' or ``representation,'' since those were terms we wanted participants to define themselves during the workshops.

\textbf{Workshop.}
To facilitate collective reflection and discussion on cultural representation in AI images, we convened four-hour workshops across three sites: two in-person workshops in Pakistan and Sri Lanka, and two online workshops in India where logistical constraints necessitated a virtual format. Online workshops were held via Google Meet.

Each workshop lasted roughly four hours and had three sections. First, participants engaged in a reflective discussion on their representational goals, drawing from the image-prompt pairs they had evaluated before the workshop. Next, they participated in prompting exercises, iteratively refining prompts through multi-turn interactions with a generative AI model capable of text-to-text, image-to-text, image-to-image, and text-to-image tasks. Finally, participants shared their experiences with image generation and representation in an evaluation discussion.

The pre-workshop evaluation task complemented the workshop by providing participants with space for individual reflection, to prepare for the group discussions. This combination allowed us to explore both individual interpretations and collective understandings of cultural representation, drawing on participants' role as experts in their cultural contexts. Individual evaluations captured participants' diverse perspectives on AI-generated cultural representation, while the workshops encouraged dialogue to uncover collective interpretations and contestations in meaning-making.\looseness=-1

\subsection{Data Analysis}
All workshops were recorded and transcribed, and where necessary, translated by authors or  local partners. All four authors participated in data preparation (e.g., data cleaning and transcribing) and analysis. We took a reflexive thematic analysis approach to analyze the workshop data, following \citet{braun2006using, braun2021one}---this includes the transcripts as well as participants' responses to the individual pre-workshop evaluation task. We met regularly as a group throughout the analysis period to inductively generate themes that captured patterns of shared meaning across workshop sections and workshops \cite{braun2021one}. We used the digital whiteboard Mural to iteratively cluster the codes into larger themes, discussing the relationship between codes and themes as we went and resolving any disagreements in synchronous group discussions. To attribute the quote we used in the following section to the participants, we have used anonymized participant identifiers: country code (IN for India, PK for Pakistan, and LK for Sri Lanka) concatenated with  a number. 

%% file: Sections/04_FINDINGS_new.tex
 We first show the granular, multi-dimensional categories of cultural representation people draw on when evaluating images, complicating the singular constructs of representation in thin evaluations. 
 We then explore how participants' goals for appropriate cultural representation were deeply situated within their specific cultural contexts, dynamically change over time, and were discursively constructed through collective dialogues. Finally, we examine the situated forms of social knowledge that underpinned these evaluations, emphasizing the importance of thick evaluations that draw on people's lived experiences and cultural expertise.

\subsection{People evaluate cultural representation through multi-dimensional categories} 
\label{findings_evaluation}

When asked to evaluate cultural representativeness of AI-generated images, our participants drew on multiple dimensions of representation to evaluate, demonstrating the need for more fine-grained categories of cultural representation than singular constructs like accuracy of representation. We distill these into five dimensions, which point to the richer categories that thick evaluations could evaluate for: \textit{incorrectness} (the accuracy of depicted physical objects---the closest to existing approaches \cite[e.g.,][]{zhang2024partiality}), \textit{missingness} (absence of iconic and expected cultural elements), \textit{specificity} (whether the subject of the image was specific to their particular sub-culture), \textit{coherence} (whether all elements of an image were appropriate given cultural and social norms), and \textit{connotation} (the symbolic meanings and interpretations associated with an image). These expanded categories of cultural representation demonstrate the need for constructs that go beyond the physical world to capture the social worlds of a culture.\looseness=-1

\subsubsection {Incorrectness} Participants evaluated the (in)correct representation of specific artifacts that have a singular corresponding physical existence outside the image, such as a building, a landmark, or a local landscape. This category has the most congruence with existing thin evaluations, as it seeks to evaluate depictions of artifacts with respect to their existence, and can often be answered with a binary or quantifiable metric since the `ground truth' is more easily determinable than the following dimensions of representation. For instance, participants talked about how the depiction of the Lotus Tower in Sri Lanka can be incorrect or correct since there is only one Lotus Tower with a very particular form. As such, when attempting to generate images, Sri Lankan participants reflected on how they were not able to produce correct images of the Lotus Tower, a famous landmark and the tallest structure in Sri Lanka. However while incorrectness was relevant for evaluating objects with clear counterparts in the physical world, it was not used to assess other forms of cultural representation of the social world, indicating that evaluations of cultural representation encompass more than just factual correctness.

\subsubsection{Missingness}
Similarly, participants evaluated whether images lacked the expected cultural elements that they felt contributed to a place's identity---what we refer to as missingness \cite[see prior work on the related, though distinct, topic of ``erasure'':][]{qadri2023,shelby2023sociotechnical,katzman2023taxonomizing,qadri2025risks}. 
Participants felt that the absence of common features and structures in images meant to represent their culture meant that a given culture or location's essence or identity would not be communicated. LK-5, a participant from Sri Lanka, noted scenes that are commonplace in Sri Lanka that were curiously missing across a number of generated images: \textit{``iconic places were never represented, like post offices and railway stations. Food items in Sri Lanka were never represented. Beaches were never reflected in any images, [but] we are an island.''} In a similar vein, LK-11 pointed out that even nationally significant elements of the Sri Lankan landscape---such as visually distinct species of trees---were absent, but should be represented: \textit{``so it should reflect more of our national items if we were to ask it to generate things about Sri Lanka, such as the Naa tree, Bo tree, national bird, animals.''} This sense of missingness was closely tied to participants' ability to distinguish one place from another and capture its unique cultural character---i.e., its iconicity \cite[cf.][]{barasch1992icon}.

 \subsubsection{Specificity} 
Participants evaluated the specificity of cultural representation---i.e., whether the depiction of cultural elements was specific to their contexts---especially for elements indicative of intersectional identities or subcultures, or cases where cultural artifacts were shared across cultures (but manifested in slightly different ways). For example, when evaluating images of women wearing \textit{saris}, participants identified the regional nuances of \textit{saris} that make them specific to their culture: \textit{``I would say it's Marathi. A Marathi sari is a little different. So, the saris also have very different variations. Right. So these saris, particularly, make me feel like it's from the Northern region. Like that is more, I think, mainstream India''} (IN-3). Evaluating for cultural specificity meant that participants paid close attention to nuanced details that distinguished the cultural practices of their social worlds from others. For other participants, when evaluating cultural representation, it was important to point out that a style of dress was more common in another part of the world than theirs: \textit{``this has more Arab cultural influence, like wearing long dresses and wearing hijabs''} (PK-8). Similar to missingness, evaluations of specificity were crucial for accurately conveying the unique aspects of a particular culture, bringing with them layers of meaning and symbolism from their social worlds.

\subsubsection{Coherence}
  
  Participants also evaluated the coherence of the images of their culture---i.e., considering the extent to which various elements in an image were in alignment about which culture they were representing and in what ways. They evaluated (in)coherence in multiple ways, such as evaluating images for unrealistic combinations of cultural elements, misaligned behaviors with social and cultural norms, and anachronistic elements of their culture. Participants identified images that contained a mishmash of cultural signifiers that did not typically occur together, such as merging cultural details from different (sub)regions, subcultures, religions, or nationalities into one image: \textit{``So the attire is of the people of Tamil. And the lanterns look Chinese, and [there is] no symbol of Sinhal[ese] new year''} (LK-10).  Other participants in this workshop elaborated on the incoherence, pointing out that:
 \textit{``the kids are wearing what normally Sri Lankan kids [are] wearing. But she is also having the [bindhi], and the elder person is kind of Sri Lankan, the mom is kind of European again. And the other kid is having totally different [clothing]---a sari on a male kid I think? So I kind of gave it a [score of] two out of five.''} (LK-1) \looseness=-1

(In)coherence was also about (mis)alignments of the perceived social behaviour in an image, where people were behaving or engaging in activities that did not align with participants' own experiences or perceptions of what was considered appropriate or logical to do in their cultures, given local social and cultural norms:  
\begin{itemize}
    \item \textit{``I've never seen women wearing jewelry in [the] Press Club. Women go there for protest, [but they are] looking like they are here for a picnic.''} [PK-8]
    \item \textit{``Even the second picture, that is also wrong. That is shown like people are boating in that place. It is never done in a religious place like a gurudwara. This is never done.''} [IN-5]
    \item \textit{``Most people go to such tombs for tourism, but it seems there's a religious ceremony or speech or proselytizing happening here in these images [like] `tableegh,' which does not happen in these tombs...''} [PK-9]
\end{itemize}

Another form of (in)coherence was temporal, when elements from different time periods were placed together, creating anachronisms. IN-5 noted that some elements of images were correct once, but no longer: \textit{``No one covers their head [now]. My mother used to cover her head. My granny used to cover her head, but I have not seen any woman who's living in a village or in a rural area [do that] nowadays. It was a part of Punjabi culture, but not anymore.''}\looseness=-1

\subsubsection{Connotations}
Finally, participants evaluated the connotations evoked by the images, recognizing that representation in visual media is not just depiction, but also a communicative act. While certain elements of images may be accurate in the sense of potentially occurring in the physical world, participants discussed how those images evoked connotations of the social world that they felt were not appropriate representations of their culture. In the workshop in Pakistan, participants noted that depictions of beards in Pakistani images evoked Western stereotypes about Pakistanis, and they raised concerns about promoting narrow ideas of Pakistani culture: \textit{``Everyone has beards in these photos. Many people in Pakistan don't have a beard. Even in this [workshop] group... this brother does not have a beard, even my beard is so small''} (PK-11). Similarly, participants interpreted connotations of poverty from the representation of particular cultural artifacts, which were not inherent to the denotations of the visual images. For example, showing women in \textit{saris} in India was associated with \textit{``a typical traditional thinking about the Indian women since long ancient times''} (IN-4). In some evaluations, participants explicitly expressed frustration as to these images being a signal of `them' seeing `us' in a particular light: \textit{``In none of the pictures we see women. I don't know what they think of us. Like we are from the 19th century and live 200 years ago''} (PK-11).\looseness=-1

\subsection{Goals for cultural representation are situated, dynamic, and negotiated}
 
\label{findings_construct}
\subsubsection{Goals for representation are situated and dynamic.}
The previous section showed the more granular categories needed to evaluate the construct of cultural representation in AI images. In this section, we identify how the goals for evaluating those constructs of cultural representation are developed through a dynamic conversation among the participants about their lived experiences and broader messaging about their culture. Participants emphasized that there was no `objective' ideal for representation, but instead what constitutes meaningful cultural representation varies both  across and within cultures and contexts. For instance,  diversity of representation is one example of a representational goal generative models might aim for, but one participant noted how identities considered diverse in other contexts were, in fact, the dominant identities in her own: \textit{``what might be diversity in the First World might actually be monotony in my area... what is diversity in a First World context or in the rest of India is the dominant culture where I come from''} (IN-15).  For IN-15, when the model attempted to achieve a decontextualized form of diversity from a Western lens, it was inadvertently just depicting dominant cultures: \textit{``In the First World, there are attempts at sort of including Muslim representations and all of that... But for example when I give a prompt as a Kashmiri woman, Muslims are the majority. Here [in Kashmir] diversity would look like something else... So here I don't want just women with hijabs represented. I also want Muslims without hijabs and non-Muslim women represented''} (IN-15).  Thus, while representational diversity was important for participants, there was no one-size-fits all approach to diversity, as it needed to be contextualized within the social worlds of the users to be meaningful.\looseness=-1

Representational goals dynamically shifted not just across countries but also within a country. For instance, IN10 from India stated the difficulty of evaluating whether the model had successfully  represented a holiday, because people celebrated the holiday so differently in different regions: \textit{``Even in the South, Ganesh chaturthi is celebrated, but the scale of it is very different. So if you look at Mumbai has very massive celebration during Ganesh chaturthi. I don't think (the celebration in) Bangalore is as big and I think some families would do a quiet celebration''} (IN-10). Participants also highlighted how even for the same person, visual representations of their own behavior would have to vary based on differences in the context they were in---such as, for instance, urban and rural, public and private, and variance in their individual adherence to cultural norms. 
To exemplify this, participant IN-3 mentioned how their attire would change if they were in rural India vs. urban India, saying \textit{``I would be wearing this in my village, yes [but not in the city]''} (IN-3). Similarly, looking at a generated image of a woman in traditional Pakistani clothes, PK-7 noted that what a woman in Pakistan might wear in the bazaar would be different from what she wore elsewhere: \textit{``We do see women around us wearing Western clothes, but in the bazaar you would see [women] dressed like this [in traditional clothes], so I don't expect women [in generated images] wearing a crop top, even though that is very me.''}

Social worlds themselves, and thus their ideal representations, also dynamically change over time, constantly evolving in ways that are not as easy to evaluate as changes to the purely physical world.  Participants noted that some aspects of cultural representation that models had generated could have been considered representative at one time, but were no longer the case, as the social world it was representing was evolving: \textit{``[This] seems like Pakistan from the 70s. Pakistan has evolved. This is a very old Pakistan. We have a Western touch now also''} (PK-11). These issues with representation across time included attire that participants thought was outdated, architectural and building styles that felt like they were from another time, and modes of transport that did not exist anymore. 

\subsubsection{Goals for representation are constructed through discursive negotiation.}
Participants' goals for and understanding of cultural representation was actively negotiated in conversation with broader social narratives they had encountered in other visual media and through dialogue with other participants during the workshop. Even seemingly objective judgments of (in)correctness were contextualized by participants within discourses and messages they had encountered outside the specific image at hand. Some participants developed goals for representation in response to stereotypes about their cultures in the media---such as perceptions that South Asian cultures were undeveloped (e.g., PK-11), leading them to want generated images that showed more modernity.
Or, when evaluating the failure of models to generate important landmarks from their cultures, participants were contextualizing these failures within perceived general power relations between the Global North and Global South that they had experienced.\looseness=-1 

Such representational goals were also reflected upon and negotiated in dialogue with other participants. One example of such negotiation was a debate that played out in one workshop on the tensions between combating stereotypes with what one participant referred to as `very clean' positive depictions compared to what another participant termed as `realism.' One participant asked their group about the desire for positive representations of their cityscapes: \textit{``What do we expect out of AI? Do we want [the images] to wash away our sins? Not have electricity poles? Do we want [the images] to be very clean?''} (PK-4). This participant then went on to say they would not want a positive representation at all times because it would not be an accurate image. One participant echoed this conclusion, noting that \textit{``there is a tension between creativity and artistic expression and realism. If you ask it to be too real, you are putting restrictions... on AI artistic capability'' }(PK-2).\looseness=-1

This tension between goals for representations to combat stereotypes 
and realism emerged concretely in images that were interpreted as depicting a less modern version of their culture. For instance, in generated images of Pakistani women, some participants noted that women wearing traditional clothes and not more `Western clothes' would convey the impression that Pakistan was not a modern place. However, other participants felt this image still reflected some women, even if it did not represent everyone, and thus was important to retain. For instance,  PK-7 argued \textit{``You have to represent the culture as opposed to a small minority of women.'' } To which PK-11 responded, \textit{``But in all pictures [of Pakistani women] we are seeing the same culture. One of the four images could be of a woman from a modern society.''} Participants also acknowledged that  there may not be consensus on what constituted desired cultural representation, 
but instead, people's goals for representation may be based on the communicative intentions of the group being asked. For instance, PK-1 noted that \textit{``country officials or ambassadors... want the best images of Pakistan. They will want [images to] be whitewashed, but we want to have stray cats and [electricity] poles''} (PK-1).\looseness=-1

While not every dialogue reached consensus, the act of dialogue helped shape their respective goals for representation. Participants noted  the importance of creating spaces for discursive engagement in the evaluation of cultural representation, allowing for the emergence of diverse perspectives and the co-construction of meaning. As one participant noted, \textit{``If you gave the same image to all five of us, we would be pointing out different elements of it... So, [in] a group discussion, you can do it more fruitfully than an individual even with a guideline. And when we move on to the next time, we will all have a more keen eye on it'' }(LK-11). This suggests that approaches to evaluating cultural representation should engage with the discursively constructed and negotiated goals for what appropriate representation looks like.

\subsection{People draw on situated knowledge and experiences to evaluate representation}
\label{findings_knowledge}

In this section, we examine the specific types of knowledge participants drew upon for different evaluative categories outlined in the first section on multi-dimensional categories. 
As participants  moved from evaluating more empirically demonstrable aspects of the physical world (e.g., is this the Lotus Tower?) to more culturally-situated judgments of the relationship between the physical and social worlds (e.g., is this scene a positive representative of my city?), participants relied on diverse forms of knowledge and deep experience of their social worlds.  For instance, evaluations of incorrectness required less cultural and social knowledge, becoming almost a form of pattern recognition. However, other forms of evaluation such as coherence or connotation required more specific knowledge, since they sought to evaluate not just thin concepts of physicality but thicker concepts of sociality. Understanding where social realities were (in)coherently displayed needed an understanding of appropriate behaviors and cultural norms that are often unstated (e.g., what lakes you would boat in and which you would not, or which generation of women might not be covering their head anymore). Even when visual elements of the physical world were being evaluated, they were often linked to social cues relying on hyper-specific visual elements: from the specific type of jewelry you would wear at particular celebrations to recognizing indigenous trees, local hairstyles, how people sit in a particular space, or even the birds that fly above a particular mosque and the potholes on a street.\looseness=-1

Participants themselves explicitly underscored the need for cultural knowledge gained through lived experience within particular social worlds to be able to evaluate the representation of different elements of those social worlds.   Participants noted that, given the rich and complex relationship between cultural artifacts and social worlds, it would be difficult for foreigners to understand the connotations or relationship between specific types of jewelry, clothing, food and cultural traditions like religious festivals. 
This knowledge was also often innate and implicit, and was not necessarily able to be made explicit in an evaluation rubric. 
When prompted to explain why a certain representation was wrong or problematic or unsatisfactory, people struggled to articulate exactly why what they were seeing was inaccurate or contextually inappropriate, saying expressions like, \textit{``there's an energy'' }(PK-7) or \textit{``there's a vibe''} (IN-3, IN-14, LK-1). Participants also noted that tourists or other outsiders may have a preconceived notion of their countries that don't match the local reality. They pointed out that relying on people to evaluate cultural representation of images from cultures other than their own would lead to worse evaluations, as outsiders might fail to recognize harmful stereotypes embedded in images, instead mistaking them for accurate portrayals.\looseness=-1 

This recognition of outsiders' limitations also prompted participants to reflect on their own inability to judge the same axes of representation for other cultures. \textit{``If someone asks me to evaluate Africa, I already have a bias[ed view] of Africa and I will evaluate accordingly''} (PK-11). While participants emphasized the importance of lived experience for accurate cultural representation, they also recognized the need for this experience to be granular and specific to intra-country social worlds. Even being from the same country, participants admitted that their personal experiences with other subcultures within their own country were often limited. Many shared examples of cultural norms and knowledge that they were not previously familiar with before this workshop. For instance, in the Pakistan workshop, participants discovered different ways of celebrating the holiday of Nauroz, surprised at their own discoveries in conversation with each other:\looseness=-1   

\begin{quote}

 PK-10: \textit{In Skardu we celebrate it differently. We have contests with eggs. The one whose egg breaks loses. We have 10 day long celebrations.} 
 
 PK-6: \textit{We celebrate it very differently. We have people gathering, people playing music, playing sports. We have different instruments. Like tambourine or rubab.}

 PK-7: \textit{I've never heard of Nauroz in Lahore or Karachi.}
\end{quote}

 These findings highlight the deeply situated cultural and social knowledge needed for evaluating appropriate cultural representation of social worlds in generated images.

%% file: Sections/05_Discussion.tex
In this discussion section,  we reflect on three implications of our findings for the practice of AI evaluations: (1) the need to reconsider the construct  of representation underpinning AI evaluations by creating congruence between that construct and its measurement, (2) the importance of co-constructing methods for evaluating representation with communities, and (3) the tensions inherent in attempts to ``thicken'' evaluation practices for cultural representation in AI.

\subsection{Thick Methods for Thick Constructs of Cultural Representation}
Our findings highlight the urgent need for a critical re-appraisal of how we conceptualize and operationalize representation in AI evaluations. Instead of representation being a singular fixed concept or objective truth, participants interpreted it in situated and negotiated ways via multiple granular categories. 
Our participants empirically demonstrated that what gets bundled into the `suitcase word' of representation \cite[cf.][]{chasalow2021representativeness} are multiple unobservable constructs of cultural representation, each requiring different approaches to measurement and metrics.  Our findings thus suggest that AI evaluations typically conflate two constructs of representation---one that is thinner, which focuses on empirical accuracy of physical worlds, and one that is thicker, such as ``coherence,'' which reflects the internal social consistency and plausibility of the social world depicted in an image.\looseness=-1 

 While thin evaluations may be suitable for assessing ``thinner'' constructs like the factual incorrectness of images' correspondence with an empirical reality or ground truth (e.g., landmarks), thicker constructs, such as those related to cultural norms, social relations, and power dynamics, necessitate thick evaluations that can capture the deeper layers of social meaning within images.  We saw in our findings that as the representational category became ``thicker'' (e.g., moving from ``incorrectness'' to ``connotation''), the evaluation process became more subjective and reliant on shared cultural understanding.  Yet the field of AI is increasingly using thin metrics to evaluate thick concepts---i.e., using  metrics and benchmarks that are better suited for evaluations of the physical world to instead evaluate the representation of social worlds \cite[cf.][]{nelson2023facct,nelson2023oxford}. We thus advocate for an evaluation ecosystem that creates congruence between the construct (i.e., cultural representation, or its dimensions) and the measurement method, which complements thin evaluations with thick evaluations.  In the words of \citet{jackson2013thin}, \textit{``[T]hin description is the necessary starting point for social investigation but not nearly enough all by itself,''} or, as \citet{rylethinking} argues, \textit{``thick description is a many layered sandwich, of which only the bottom slice is catered for by the thinnest description.''}\looseness=-1

\subsection{Co-Constructing Evaluations of AI's Cultural Representation with Communities}

Our study shows that there is no objective conceptualization of representation to optimize for in images. What should ideally appear when a model is prompted for ``Indian,'' ``Pakistani,'' or ``American'' is a culturally- and contextually-contingent choice. Thus, most representational evaluations require evaluators to make interpretations, involving culturally-situated judgment, rooted in social experience.  Who gets to decide what representation is and how a culture should be depicted?  Who has the power to shape meaning in the struggle over meaning-making that is representation in media \cite[cf.][]{desai2000imaging,hall1997representation,abu1991before}? Currently, the power of this choice lies with AI researchers and anonymous evaluators who often have limited experience with the social worlds subject to evaluation. We argue, as with AI fairness more broadly, that we must \textit{``bring the people back in''} to AI measurement of representation \cite{denton2020bringing,Smith-design-fairness}, to make it more contextual and remove translation gaps between abstract metrics and the actual risks and harms they seek to measure \cite{jacobs2021measurement,d2023data}.\looseness=-1

Although some prior work has conducted qualitative evaluations of representation in AI, they have primarily focused on
on differences in stakeholders' judgments or decisions compared to some gold standard, with less focus on differences in the underlying \textit{evaluative processes} that inform those judgments. As an example, disagreement itself is a robust area of research in NLP and AI, with much focus on the sociocultural influences that shape human judgment \cite{davani2022dealing, diaz2022crowdworksheets}. Studying the culturally-situated ways that people evaluate and understand representation in our study allowed us to suggest metrics and measures of representation that are more in line with the ways people who actually experience and consume images interpret representation of their culture. Thus, our findings echo researchers like \citet{Smith-design-fairness} who call for AI measurement to come in line with the actual experiences of people on the ground---not just the experiences of researchers---and integrate their experiences, knowledge, and expertise \cite[cf.][]{diaz2024makes} in the measurement process. The aim is not simply to apply the same metrics in different ways, but to co-construct metrics and measures with stakeholders. Integration of expertise for representation must go beyond large-scale, globally-deployed surveys often used by AI researchers. This approach offers only a very limited mechanism to integrate cultural knowledge from particular communities into evaluative frameworks, because the metrics themselves are developed by AI researchers. Annotation evaluation pipelines are structured to align resulting data with what data requesters define to be ground truth  \cite{miceli2020between, posada2023platform, chandhiramowuli2024making, muller2021designing}. As a result, annotators or evaluators must constrain their assessments within the feedback categories set by researchers. Evaluations outside of these constraints for evaluation tasks are conflated with other concerns, or dismissed altogether \cite{posada2023platform}.\looseness=-1

Building on work on subjectivity in annotation \cite{cabitza2023toward, davani2022dealing} and work in HCI that contests the idea of singular ground truth \cite{muller2021designing,gordon2022jury}, our study suggests opportunities for developing evaluation methods for representation that leverage discursiveness and deliberation \cite[cf.][]{shen2022model}. \citet{bergman2024stela} have highlighted the value of discursive deliberation in eliciting individuals' justifications for the views they espouse.  To do this, evaluative methods might actively encourage collective conversation and disagreement among evaluators through interpretive methods like workshops in addition to thin methods like benchmarks. 
In our study, discursive evaluation helped to fill knowledge gaps, such as whether people in different regions celebrated certain occasions in the same way. However, our discursive evaluations also revealed and helped to resolve disagreements that were ontological in nature, about the very nature of representation. Ontological disagreement here was not simply a matter of knowledge gaps, but instead fundamental differences in understanding what an image represents and what it \textit{should} represent. Discursive evaluation processes both shed light on ontological disagreements and shift understandings in real time. Thus, dialogue and discussion can change or reaffirm evaluative judgments evaluators might have held before---changing the nature of feedback evaluators might give in individual annotation tasks.\looseness=-1

\subsection{Making Space for Thickness in AI Practice} 

As we build the case for thick evaluations for cultural representations, we also reflect on the epistemological hurdles that may emerge if thick evaluations were to be adopted by AI developers. For one, thick evaluations are at odds with incentives and values like generalization and scale in computing \cite{birhane2022values, hanna2020towards}. Thick evaluations demand a deep engagement with situated cultural contexts \cite[cf.][]{arzberger2024nothing}, requiring time and investment, and acknowledging the diversity of interpretations and meanings associated with representation. They embrace subjectivity and multiple interpretations, acknowledging that there is no single ``correct'' way to represent a social world. This all limits their scalablity and generalizability---but at the same time, they also empirically \textit{demonstrate} the limits of scale and generalizability. As \citet{tsing2012nonscalability} reminds us, the multiplicity of social worlds resists attempts at hegemonic structuring into scalable units.\looseness=-1

For representational evaluations to become `thicker,' AI practitioners and researchers will also have to stop drawing on the epistemic cultures of positivism and techno-solutionism prevalent in ML and computing  \cite{morozov2013,lindtner,cunningham}.  These epistemologies in AI   have been extensively critiqued in AI fairness. For instance,  annotation practices rarely account for social histories of race and gender \cite{scheuerman2020we}, and more broadly, algorithms fail to capture social nuances and lived realities \cite[e.g.,][]{broussard2018artificial,mattern2021,arzberger2024nothing}. The critiques of thin-ness our findings proffer are also in line with critiques of positivism in the social sciences.  For instance, \citet{babones2016} argues against treating humans and society as a \textit{``knowable objective reality, represented reasonably well by the variables''}   and \citet{kitchinbigdata} argues that positivism produces social analysis that  \textit{``is reductionist, functionalist and ignores the effects of culture, politics, policy, governance and capital.''}  In this paper, we demonstrate how current approaches to evaluating representation thus ignore extensive critiques of quantitative measurement in the social sciences, which caution against attempting to formalize fundamentally social processes, like cultural representation, into quantifiable metrics or objective functions that assume a singular empirical reality \cite{steinmetz2005introduction}. Our findings demonstrate how limited the prevailing paradigms of objectivity and quantifiability are when attempting to evaluate social worlds, and how much richer AI evaluations could be  if we embraced the pluralism of thick evaluations, with their emphasis on situated context, subjectivity, and lived experiences. Such a move would require de-centering objectivity, acknowledging the situatedness of evaluations of cultural representation.\looseness=-1

 Our findings  also show the need for a fundamental shift in how we understand and value knowledge in AI practice. The practice of thickness needs communities' participation in co-constructing evaluations. That is why in this paper we draw on interpretive, dialogue-based methods to capture the discursive processes involved in negotiations of the meaning of images. Our methods also recognized the expertise brought by local experts and stakeholders who could partner with us to create a sample situated in the diverse social worlds of participants. However, adopting this practice would require AI researchers to recognize and embrace pluralistic ways of knowing and understanding the world.  Integrating qualitative, situated knowledge into the predominantly quantitative realm of AI necessitates a shift in mental models and a confrontation with deeply ingrained assumptions about considerations of expertise \cite[cf.][]{madaio2024learning,diaz2024makes}. We take inspiration from those like Agre who call for a critical technical practice around computing \cite{agre2014toward,malik2022critical}, and those like \citet{nelson2023facct,nelson2023oxford} to call for a practice of thickness not just in measurement but also in AI writ large.  This would require questioning the singular focus on scale and generalization, embracing qualitative methods and epistemologies, actively co-creating AI with communities and reflexively interrogating the epistemological values embedded in our evaluation practices.\looseness=-1

%% file: Sections/06_Limitations.tex
This work has several limitations, which future work may address. First, our participants were from three countries in South Asia; bringing their own cultural and individual perspectives on the most salient elements for which to evaluate representation. Since generalizability  is not the goal of interpretive qualitative research, which instead aims for \textit{transferability}, \cite{soden2024evaluating,drisko2025transferability}, we do not make any claims of generalizability of this research to other locales. Thus, future research could help shed light on the extent to which similar taxonomies of thick evaluations  may be transferrable \cite{drisko2025transferability} or extendable for representational evaluations with different contexts or populations. In addition, we focused on text-to-image AI systems in this study. Although we believe our work could speak to thick evaluations of representation in other AI modalities (e.g., text, video, audio, or multi-modal systems), it is an open question of precisely how such thick evaluations in other modalities ought to be conducted, or how the approach ought to change, if so. In this work, we discuss how certain elements of cultural representation may be best evaluated by thick evaluations, while others (e.g., the ``incorrectness'' dimension) may be better suited to thin evaluations, or correspondence to a ground truth; however, future work should be conducted to validate this hypothesis, to better understand which elements of cultural representation are best evaluated with which approaches. Finally, although in this paper, we map out a space for conducting thick evaluations, future work should explore how best to incorporate the results of these evaluations into shaping the design and development of generative AI systems, in order to avoid the failures of cultural representation that motivated this work.\looseness=-1 

%% file: Sections/07_Conclusion.tex
 
 As AI technologies are being proposed to engage in cultural production, and research continues to highlight the failures of AI to adequately represent all cultures, our work suggests the need to develop  new, thicker forms of evaluations of cultural representation, to better reflect how people consume and interpret images of their cultures. Thus, in this paper, we introduce a framework for thick evaluations of cultural representation, developed in conversation with participants in South Asia, that can move the field beyond existing reductive ideals of representation. Showing that representation is a process of meaning-making through the interaction between image, viewer, and context, we argue that measurement of representation in AI needs to encompasses more than just factual correctness, creating practices that bridge the disconnect between constructing representation as a technical goal and understanding it as a social goal. Developing congruence between the thickness of representation as a concept and thickness of the evaluation method will ensure AI practitioners avoid creating brittle, impoverished evaluations and mitigations for representation in the design of AI systems, and, ideally, lead towards AI systems that are better able to represent the plurality of social worlds of all cultures.\looseness=-1

%% file: Sections/08_Appendix.tex
\section{Appendix}

We include here a table of detailed demographic information of our participants from our study. We present their details by countries, please see the tables below:

\begin{table*}[!ht]

 \scriptsize  
    \begin{tabular}{|c|c|c|c|c|c|c|}
    \hline
        \textbf{Participants} & \textbf{Country}  & \textbf{Age}  & \textbf{Gender} & \textbf{Educational Level} & \textbf{Region}  & \textbf{Language} \\ \hline
        LK-1 & Sri Lanka & 25–34 & Male & Undergraduate  & North Western province & Sinhala \\ \hline
        LK-2 & Sri Lanka & 18--24 & Male & Undergraduate & North Western Province & Sinhala \\ \hline
        LK-3 & Sri Lanka & 18--24 & Female & Undergraduate   & Western Province & Tamil, English  \\ \hline
        LK-4 & Sri Lanka & 18--24 & Male & Secondary  & Western Province & Sinhala \\ \hline
        LK-5 & Sri Lanka & 18--24 & Female & Undergraduate & Western Province & Sinhala \\ \hline
        LK-6 & Sri Lanka & 18--24  & Female  & Undergraduate & Western Province/ Central Province & Sinhala  \\ \hline
        LK-7 & Sri Lanka & 18--24 & Female & Undergraduate & Central Province/ Western Province & Sinhala,Tamil and English \\ \hline
        LK-8 & Sri Lanka & 25–34 & Male & Undergraduate  & Western Province/ North Western Province & Tamil \\ \hline
        LK-9 & Sri Lanka & 18--24  & Male & Undergraduate & Western Province/ Southern Province/ Central province & Sinhala \\ \hline
        LK-10 & Sri Lanka & 35–44  & Male & Undergraduate & Western Province & Sinhala  \\ \hline
        LK-11 & Sri Lanka & 45–54 & Female  & Secondary & Western Province  & Sinhala  \\ \hline
    \end{tabular}
 \caption{Participant Information from Sri Lanka}

\end{table*}

\begin{table*}[!ht]
    
     \scriptsize 
    \begin{tabular}{|c|c|c|c|c|c|c|}
    \hline
        \textbf{Participants} & \textbf{Country}  & \textbf{Age}  & \textbf{Gender} & \textbf{Educational Level} & \textbf{Region}  & \textbf{Language} \\ \hline
        PK-1 & Pakistan & 25–34 & Female & Post-Graduate  & Punjab  & Urdu  \\ \hline
        PK-2 & Pakistan & 25–34  & Male & Undergraduate & Gilgit-Baltistan  & Balti \\ \hline
        PK-3 & Pakistan & 25–34 & Male & Post-Graduate  & Rajanpur & Balochi, Siraiki, Urdu \\ \hline
        PK-4 & Pakistan & 35–44 & Transgender Man & Undergraduate & Punjab & Urdu \\ \hline
        PK-5 & Pakistan & 18--24 & Transgender woman & Undergraduate  & Punjab & Punjabi, Urdu \\ \hline
        PK-6 & Pakistan & 18--24 & Female & Undergraduate  & Gilgit  & Burushaski \\ \hline
        PK-7 & Pakistan & 18--24 & Female & Secondary & Punjab  & Urdu, Punjabi \\ \hline
        PK-8 & Pakistan & 18--24 & Female & Undergraduate  & Punjab & Urdu, Punjabi \\ \hline
        PK-9 & Pakistan & 18--24 & Male & Secondary & Gilgit Baltistan  & Balti \\ \hline
        PK-10 & Pakistan & 18--24 & Male & Secondary  & Gilgit Baltistan & Balti \\ \hline
        PK-11 & Pakistan & 18--24 & Male & Undergraduate  & South Punjab & Saraiki \\ \hline
    \end{tabular}
\caption{Participant Information from Pakistan}
\end{table*}

\begin{table*}[!ht]
    \scriptsize
    
    \begin{tabular}{|c|c|c|c|c|c|c|}
    \hline
        \textbf{Participants} & \textbf{Country}  & \textbf{Age}  & \textbf{Gender} & \textbf{Educational Level} & \textbf{Region}  & \textbf{Language} \\ \hline
        IN-1 & India & 25–34 & Female & Post-Graduate & Kerala & Malayalam \\ \hline
        IN-2 & India & 55–64 & Female & Undergraduate & Maharastra/ Karnataka  & English, French, Hindi \\ \hline
        IN-3 & India & 25–34 & Female & Undergraduate & Assam & Assamese and Hindi \\ \hline
        IN-4 & India & 45–54 & Female & Post-Graduate & Maharashtra & Marathi \\ \hline
        IN-5 & India & 45–54  & Female & Post-Graduate& Chandigarh/ Punjab & Punjabi, Hindi, English \\ \hline
        IN-6 & India & 25–34 & Male & Post-Graduate & Bihar & Hindi \\ \hline
        IN-7 & India & 18--24 & Male & Secondary & Gujarat  & Gujarati \\ \hline
        IN-8 & India & 25–34  & Non-binary & Post-Graduate & Maharashtra & Hindi, Magadhi \\ \hline
        IN-9 & India & 25–34 & Genderqueer & Post-Graduate & West Bengal, Gujarat, Karnataka & Bengali and English  \\ \hline
        IN-10 & India & 25–34 & Female & Undergraduate & Telegana/ Karnataka & Telugu, Hindi, Kannada \\ \hline
        IN-11 & India & 18--24 & Male & Undergraduate & Rajasthan & Rajasthani \\ \hline
        IN-12 & India & 18--24 & Male & Undergraduate & Chhattisgarh & Chhattisgarhi\footnote{Chhattisgarh local Language mix with Hindi known as Chhattisgarhi} \\ \hline
        IN-13 & India & 25–34 & Male & Post-Graduate & Tripura & Sylheti  \\ \hline
        IN-14 & India & 25–34 & Female  & Post-Graduate & Assam & Assamese \\ \hline
        IN-15 & India & 35–44 & Female & Post-Graduate & Kashmir & Kashmiri, Urdu, Hindi, English \\ \hline
    \end{tabular}
    \caption{Participant Information from India}
\end{table*}



